\documentclass[journal=jacs,manuscript=communication,layout=twocolumn]{achemso}
\setkeys{acs}{articletitle = true}
\setkeys{acs}{maxauthors = 0}
\pdfoutput=1

\usepackage[T1]{fontenc}
\usepackage{multirow}
\usepackage{color}
\usepackage{xspace}

\newcommand{\sub}[1]{\ensuremath{_{\mathrm{#1}}}}

\newcommand{\Angst}{\AA\xspace}

\author{Ravishankar Sundararaman}
\affiliation{Department of Materials Science and Engineering, Rensselaer Polytechnic Institute, Troy, NY 12189, USA}
\email{sundar@rpi.edu}

\author{Marta C. Figueiredo}
\affiliation{University of Copenhagen Department of Chemistry Nano-Science Center Universitetsparken, 5 2100 Copenhagen, Denmark}

\author{Marc T. M. Koper}
\affiliation{Leiden Institute of Chemistry, Leiden University, PO Box 9502, 2300 RA Leiden, The Netherlands}

\author{Kathleen A. Schwarz}
\email{kas4@nist.gov}
\phone{(301) 9752821}
\affiliation{Material Measurement Laboratory, National Institute of Standards and Technology, Gaithersburg, MD 20899, USA}

\title{Electrochemical Capacitance of CO-terminated Pt(111) is Dominated by CO-Solvent Gap}

\abbreviations{DFT}

\let\oldmaketitle\maketitle
\let\maketitle\relax

\begin{document}

\twocolumn[\begin{@twocolumnfalse}
\oldmaketitle
\noindent\begin{minipage}{4.7in}
\begin{abstract}
The distribution of electric fields within the electrochemical double layer
depends on both the electrode and electrolyte in complex ways.
These fields strongly influence chemical dynamics in the
electrode-electrolyte interface, but cannot be measured directly with sub-molecular resolution.
We report experimental capacitance measurements for aqueous interfaces of CO-terminated Pt(111).
By comparing these measurements with first-principles density-functional theory (DFT) calculations,
we infer microscopic field distributions and decompose contributions
to the inverse capacitance from various spatial regions of the interface.
We find that the CO is strongly electronically coupled to the
Pt, and that most of the interfacial potential difference
appears across the gap between the terminating O and water,
and not across the CO molecule as previously hypothesized.
This `gap capacitance' resulting from hydrophobic termination lowers
the overall capacitance of the aqueous Pt-CO interface, and makes it
less sensitive to electrolyte concentration compared to the bare metal.
\end{abstract}
\end{minipage}\hspace{0.3in}\begin{minipage}{2.5in}{\includegraphics[width=2in]{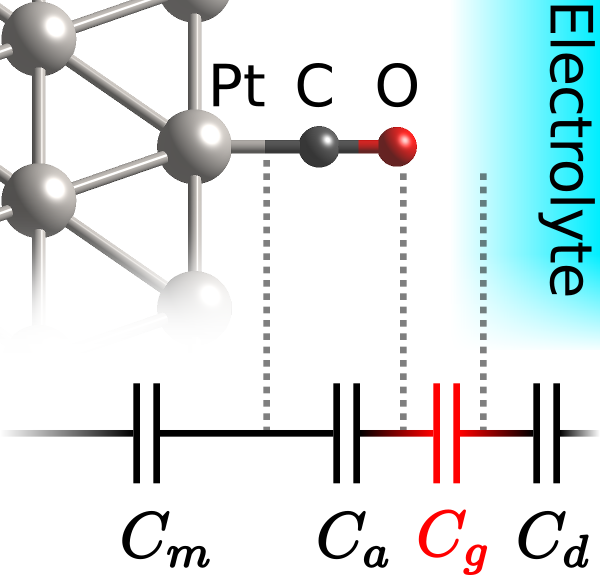}}\end{minipage}
\end{@twocolumnfalse}]

Chemical processes at electrochemical interfaces are fundamentally important
for a wide range of technological applications including chemical synthesis,
corrosion prevention, energy conversion and energy storage.~\cite{taylor2,Hersam,CopperCO2,Chan,COreductionCu111}
The mechanisms, kinetics and selectivity of these processes depend critically
on the microscopic structure of the electrode-electrolyte interface.
The distribution of electric fields and potentials at this interface
is particularly important because it determines the adsorption/desorption
energetics of relevant chemical species and their charge states at the surface.~\cite{Chan,COreductionCu111, Schwarz2015}
The basis for understanding field distributions in electrochemical interfaces
is the double layer model of electrochemical capacitance.~\cite{Bard}
The total capacitance decomposes into two capacitors in series:
an `inner layer' or Stern capacitance dependent on electrode-solvent-ion interactions,
and the Gouy-Chapman capacitance from ionic contributions in the outer `diffuse' layer.~\cite{Stern}

The double layer model of electrochemical interfaces has been tested
extensively for bare-metal electrode surfaces,\cite{ParsonsZobel,Schmickler}
but its applicability to metal surfaces with adsorbates has
not yet been investigated as thoroughly.\cite{crispin,white, xu}
Recent experiments using Stark tuning as a probe of the local electric fields 
of adsorbates find results in support of the double layer model.\cite{Lian,jahan,jahanJACS}
Metal surfaces terminated by a strongly-bonded rigid adsorbate are
most suitable for fundamental studies of the structure of such an interface
that include first-principles calculations because the electrode+adsorbate
structure remains in a single configuration with a purely electronic response,
and does not require thermodynamic averaging over rotational and flexional modes.
CO-terminated Pt surfaces are particularly well characterized,
with electrochemical capacitance measurements in a range of solvents
including water, acetonitrile and ionic liquids.~\cite{Weaver1992,Baldelli,koper2002}
The classical double layer model for this system consists of an inner layer
with a linear potential drop covering the CO adlayer and extending till
the outer Helmholtz plane, assumed to be one radius of unsolvated
electrolyte ion beyond the first water layer.\cite{Weaver1992}
Smith and White~\cite{white} proposed an alternate model for electroactive films
in which the potential drops linearly across a dielectric, and then decays rapidly
in the electrolyte.

It is challenging to experimentally evaluate these models because the sub-molecular field
distributions within the electrode, adlayer and electrolyte cannot be measured directly in experiment.
Direct measurement of microscopic fields is possible using Stark shift spectroscopy,
but with a resolution limited to the size of the vibrational chromophore molecule added to the surface.
Additionally, the adsorbed Stark tuning probe itself changes the field distribution and the double layer.
First-principles calculations can probe field distribution over sub-molecular
length scales, with density-functional theory (DFT) providing the structure and electronic
response of the electrode + adlayer, coupled to suitable solvation models
that capture the response of the electrolyte.

In this Letter, we combine experimental capacitance measurements with solvated
DFT calculations to understand the aqueous Pt-CO interface in microscopic detail.
We find good agreement between the DFT + continuum solvent predictions
and experimental measurements for the total capacitance.
The calculations additionally provide the atomic-scale variation
of the potential across the metal, adlayer and electrolyte.
We find two approximately-linear potential regions within the inner layer:
a mostly flat potential region in the CO adlayer, and a far steeper potential
profile in the gap between the terminal O and the start of the water.
The `gap capacitance' due to this second region contributes significantly
to the lower capacitance and relative insensitivity to ionic concentrations,
compared to metal electrodes, and illustrates the limitations of the
classical inner/outer-layer division for hydrophobic adsorbates.\cite{GalliTerminations}

\begin{table}
\caption{The experimental double layer capacitances and standard deviations
for bare Pt(111) and Pt(111) with adsorbed CO, extracted from
measurements of the current density at 0.4 V at different scan rates. 
Below, `H\sub{upd} corrected' refers to correction of the capacitance
for under-potential deposition of H as described in the Supporting Information.
\label{tab:Expt}}
\centering
\begin{tabular}{|c|c|c|}
\hline
HClO$_4$ conc.  & \multicolumn{2}{c|}{Capacitance [$\mu$F/cm$^{2}$]} \\ \cline{2-3}
[mmol/L] & Pt(111) & Pt(111)/CO\sub{ads} \\
\hline
1   & 70.2$\pm$1.5 & 11.0$\pm$0.3 \\
10  & 85.2$\pm$1.1 & 10.8$\pm$0.5\\
100 & 106.0$\pm$1.9 & 11.2$\pm$0.7 \\
100 & 20 \cite{pt111Cap} (H\sub{upd} corrected) &  \\
100 & 14$\pm$5 \cite{lipkowskiCap} (H\sub{upd} corrected) &  \\
\hline
\end{tabular}
\end{table}

\begin{table*}
\caption{DFT predictions of total capacitance of aqueous bare Pt(111)
and Pt(111) with adsorbed CO, and the spatial decomposition of
the DFT capacitances and inverse capacitances (from Fig.~\ref{fig:Cinv}).
\label{tab:Cap}}
\centering
\begin{tabular}{|c|c|c|cccc|cccc|}
\hline
\multirow{3}{*}{System} & \multirow{3}{*}{Method}
& \multicolumn{5}{|c|}{Capacitance [$\mu$F/cm$^2$]}
& \multicolumn{4}{|c|}{Inverse Capacitance [cm$^2$/$\mu$F]} \\
\cline{3-11}
& & \multirow{2}{*}{Total} & Pt & CO & Gap & Diffuse & Pt & CO & Gap & Diffuse \\
& & & ($C_m$) & ($C_a$) & ($C_g$) & ($C_d$)	& ($C_m^{-1}$) & ($C_a^{-1}$) & ($C_g^{-1}$) & ($C_d^{-1}$) \\
\hline
\multirow{2}{*}{Pt}
&  GGA & 27.5 & 784. & - & 32.0 & 259. & 0.0013 & - & 0.0312 & 0.0039 \\
&  LDA & 26.4 & 598. & - & 31.0 & 250. & 0.0017 & - & 0.0323 & 0.0040 \\
\hline
\multirow{2}{*}{Pt-CO}
&  GGA & 9.7 & 228. & 39.5 & 14.3 & 252. & 0.0044 & 0.0253 & 0.0698 & 0.0040 \\
&  LDA & 9.4 & 185. & 39.2 & 14.0 & 235. & 0.0054 & 0.0255 & 0.0716 & 0.0043 \\
\hline
\end{tabular}
\end{table*}

The voltammetric experiments were performed in a three-electrode configuration using a Pt (111) bead (from icryst, www.icryst.com)
 as working electrode, a Pt wire as counter and a reversible hydrogen (RHE) as reference electrode, at room temperature. 
[Note:  Certain commercial materials are identified in this paper to foster understanding. Such identification does not imply 
recommendation or endorsement by the National Institute of Standards and Technology, nor does it imply that the materials
 or equipment identified are necessarily the best available for the purpose.]
The electrolyte solutions were prepared using different concentrations of HClO$_4$ (70 \%, Merck Suprapur) and ultrapure water
 (Merck Millipore, 18.2 M$\Omega$ cm). The electrochemical measurements were performed with the working electrode in hanging
 meniscus configuration and the potential was controlled with an Autolab PGSTAT302N potentiostat. The current density reported
 represents the measured current normalized to the electrochemical surface area of the working electrode. To ensure the 
proper surface ordering, the electrodes were prepared as previously described\cite{expt1,expt2}.  Briefly, prior to each measurement, 
the crystals were flame-annealed and cooled to room temperature in an Ar:H2 (3:1) environment. Subsequently, the crystal 
was protected with a drop of water saturated in the same gas mixture and transferred to the electrochemical cell. All the
 experiments were performed by first acquiring a blank voltammogram of the Pt(111) in the electrolyte solution purged with
 Ar (6.0 Linde), to ensure the surface cleanliness and order. After recording the blank, CO (6.0 Linde) was purged in the 
solution for 2 minutes to ensure a full layer of CO on the electrode surface, and subsequently Ar was purged for 20 min 
through the electrolyte solution in order to remove all the CO from solution.
During this process the electrode potential was kept at 0.1 V (vs RHE) to avoid CO oxidation. 
We only consider full CO coverage in the work as it is most reproducible experimentally
and amenable for theoretical analysis of capacitance contributions due to its planar homogeneity.
See Supporting Information for experimental voltammograms and
details on the extraction of capacitance from the measurements.

Computationally, we follow the protocol previously established in Ref.~\citenum{SolvEval}.
Briefly, we perform plane-wave electronic DFT calculations in the JDFTx code,\cite{JDFTx}
with the Perdew-Burke-Ernzerhof (PBE) exchange-correlation functional,\cite{PBE}
ultrasoft pseudopotentials,\cite{GBRV} and plane-wave kinetic energy cutoffs
of 20~$E_h$ (Hartrees) and 100~$E_h$ for orbitals and charge densities respectively.
We treat the bare and CO-terminated platinum surfaces with inversion-symmetric
slabs with five Pt(111) layers, and use truncated Coulomb potentials\cite{TruncatedEXX}
to eliminate interactions with periodic images normal to the slab.
We use a nonlinear continuum solvation model\cite{NonlinearPCM}
that captures nonlinearities in both the dielectric response of the solvent
and ionic response of the electrolyte within a local-response approximation.
We employ the original parametrization of this model based on solvation energies
of organic solutes for the CO-terminated surface calculations, and the revised 
parametrization for metallic surfaces\cite{SolvEval} for the bare surface calculations.
To evaluate capacitances and potential profiles, we calculate the change
in electron number and electrostatic potential\cite{ElectrostaticPotential}
between a neutral DFT calculation and a grand canonical DFT calculation\cite{GC-DFT}
at a potential fixed 0.1~V below the neutral value.
The solvation models used here have been carefully benchmarked for
prediction of electrochemical properties,\cite{SolvEval,COreductionCu111}
while equivalent treatment with \emph{ab initio} molecular dynamics would
require impractically large simulation cells to contain statistically meaningful
numbers of ions in the electrolyte.\cite{GC-DFT}  Importantly, the key conclusions
below relate to the electronic structure of the electrode calculated in DFT,
and are therefore insensitive to the details of the solvation model.

Tables~\ref{tab:Expt} and \ref{tab:Cap} list the total capacitance from
these calculations and from the experimental measurements.
The DFT calculated capacitance reduces from 26~$\mu$F/cm$^2$ for the bare surface
to 10~$\mu$F/cm$^2$ for the CO-terminated surface.
The DFT results are insensitive to the exchange-correlation functional
and are in good agreement with experiment for the CO-terminated surface.  For the
Pt(111), we note that the experimental capacitance measurements are performed near the
H$_{upd}$ region, and include contributions from hydrogen adsorption onto the surface.
When the experimental values are corrected for the hydrogen adsorption, they are considerably
lower and closer to the values predicted from DFT calculations~\cite{pt111Cap,lipkowskiCap}.

Next, to better understand the reason for the lowered capacitance for Pt-CO,
we compare the spatial distribution of the DFT electrostatic potential
for the CO-terminated and bare surfaces.
Specifically, we define the inverse capacitance profile,
\begin{equation}
C^{-1}(z) \equiv \frac{\partial\bar{\phi}(z)}{\partial N},
\end{equation}
where $\bar{\phi}(z)$ is the electrostatic potential averaged along the directions
parallel to the surface and $N$ is the number of electrons in the DFT calculation.
We calculate the derivative from the difference between calculations
at the neutral potential $V_0$ and a potential $V_1 = V_0 - 0.1$~V.
With this definition, $C^{-1}(0)$ (with $z=0$ inside the electrode)
is the inverse of the total capacitance, and its spatial profile
elucidates the contributions of each spatial region to
the inverse capacitance (that combine in a series model).

\begin{figure}
\includegraphics[width=\columnwidth]{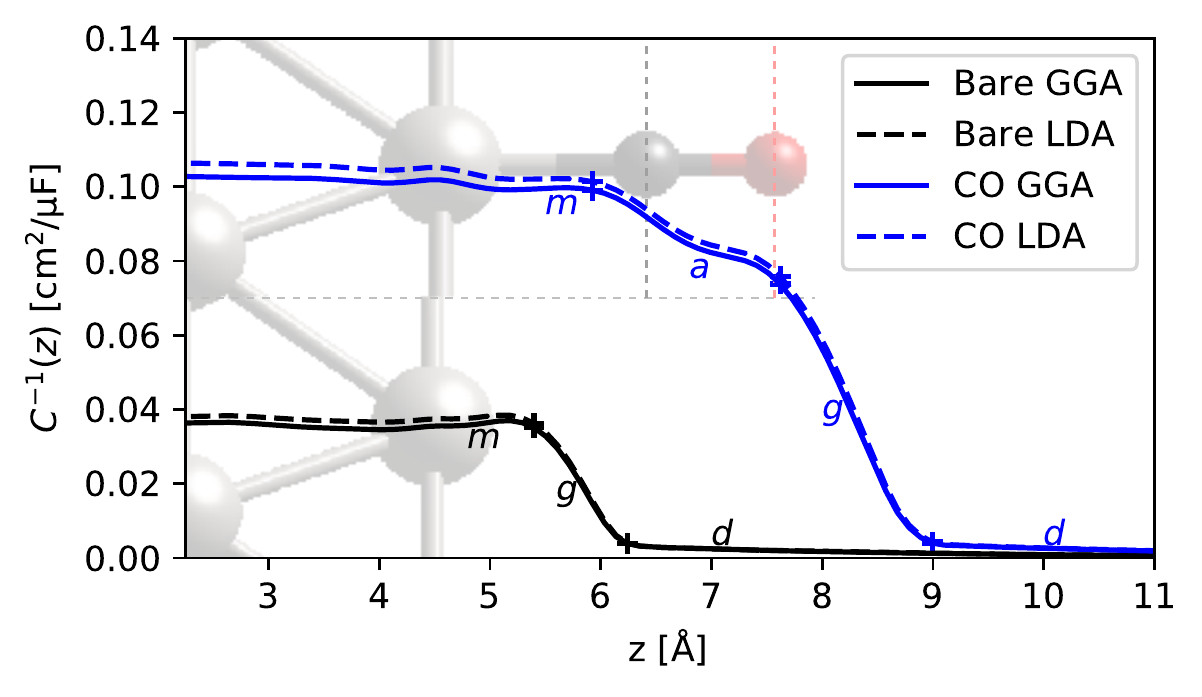}
\caption{Inverse capacitance for bare Pt(111) and Pt(111) terminated (atop) by CO.
Dotted lines indicate the locations of C (gray) and O (red),
as also shown by the atomic configuration in the background.
Note that the center of the slab is at $z=0$, and `$+$'s mark the points
of maximum curvature for separating series capacitor regions,
$m$ (metal), $a$ (adlayer, only for CO case), $g$ (gap)
and $d$ (diffuse) (see Table.~\ref{tab:Cap}).
\label{fig:Cinv}}
\end{figure}

Figure~\ref{fig:Cinv} shows the inverse capacitance profiles of the
Pt and Pt-CO interfaces, compared between GGA and LDA DFT calculations.
The potential is approximately constant inside the metal, which manifests
as a flat $C^{-1}$ profile up to the metal surface at $z \approx $5~\Angst .
The bare Pt system has a lower inverse capacitance (higher total capacitance)
in this flat region which extends till $\approx 1$~\Angst away from the surface Pt atom.
In contrast, the Pt-CO has a higher inverse capacitance (lower total capacitance),
which reduces slowly across the CO, and then drops quickly past the terminal O atom.
In both cases, most of the potential drop, and hence the $C^{-1}$ change,
occurs in the gap region between the outermost surface atom and the fluid.
The results are mostly insensitive to the DFT functional, indicating that
the spatially-resolved predictions are also relatively robust against DFT errors.

To further understand the drop in capacitance, we decompose the inverse capacitance
across the entire interface into contributions from different spatial regions.
We separate regions at the points of maximum magnitude of curvature in $C^{-1}(z)$
(points where the slope changes the most, i.e. where the induced charge density peaks),
marked with `+'s in Fig.~\ref{fig:Cinv}.
Both the bare and CO-terminated systems exhibit a metal region `$m$'
at the left end and the diffuse region of the electrolyte `$d$' at the right end.
In the bare surface, these regions are separated by the gap region `$g$',
whereas for the CO-terminated surface, there is an additional region `$a$'
corresponding to the adlayer between the $m$ and $g$ regions.
The total capacitance is given by the series capacitor model,
$C^{-1}\sub{tot} = C_m^{-1} + C_a^{-1} + C_g^{-1} + C_d^{-1}$,
where $C_a^{-1}$ is present only for the CO-terminated case.

Table~\ref{tab:Cap} shows the separation of the total
(inverse) capacitance into these spatial contributions.
For both the bare and CO-terminated systems in 1 M (mol/L) aqueous electrolyte,
the metal and diffuse inverse-capacitance contributions are negligible;
the diffuse contribution will become more important at lower ionic concentrations
near the potential of zero charge as usual within the Gouy-Chapman-Stern theory.~\cite{Bard}
Importantly, note that for the CO-terminated Pt surface, the series capacitance
is dominated by the gap region, $C_g\approx 14~\mu$F/cm$^2$,
rather than  by the adlayer, $C_a \approx 40~\mu$F/cm$^2$
(lower capacitance dominates in series).
Also note that this gap region capacitance for the CO-terminated surface
is much smaller than that of the bare surface, $C_a \approx 30~\mu$F/cm$^2$.
This suggests that the total capacitance of the Pt-CO interface is low
because the surface repels solvent, rather than because the capacitance
of the Pt-CO itself is low with the CO adlayer acting as
an insulating spacer as previously suggested.\cite{Weaver1992}

\begin{figure}
\includegraphics[width=\columnwidth]{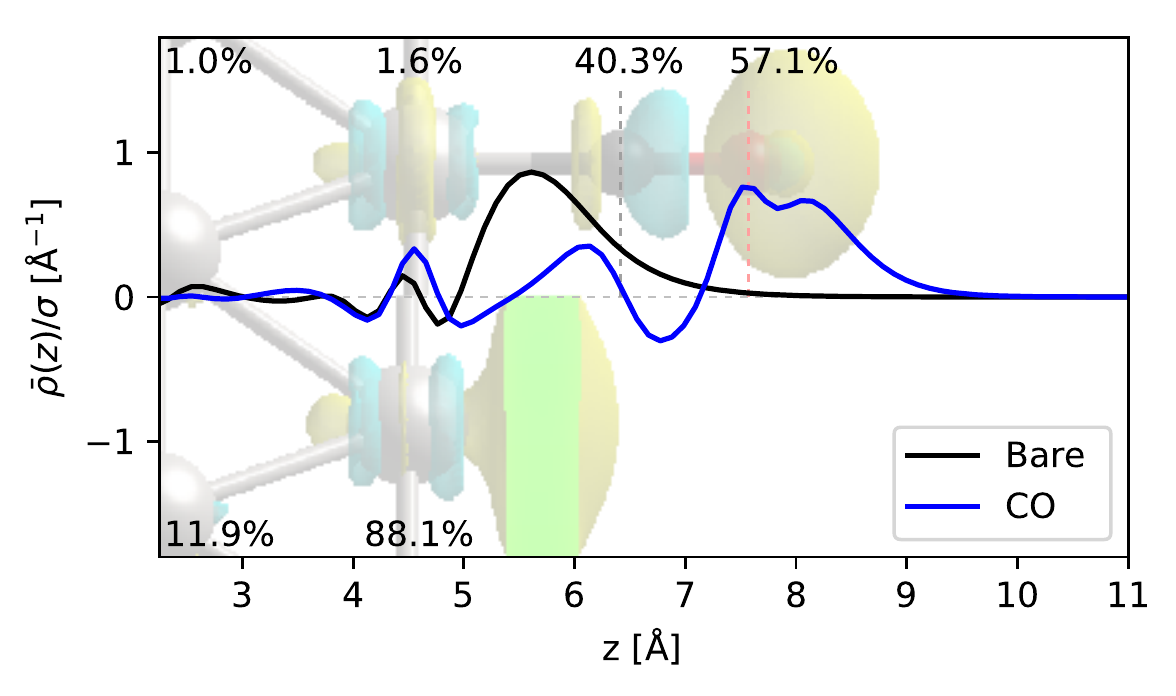}
\caption{Spatial profile of the planarly averaged induced electronic charge density in bare and CO-terminated Pt surfaces,
with an isodensity surface of the induced charge distribution shown in the background image.
The labels indicate the fraction of the induced charge in the neighborhood of each atom,
calculated using Bader analysis, to the total induced charge on the surface.}
\label{fig:Charge}
\end{figure}

In fact, the CO adlayer appears to act more like an extension of the metal,
as shown by the spatial distribution of the induced charge density in Fig.~\ref{fig:Charge}
(the planarly averaged charge density $\bar{\rho}$(z) is normalized by
the total surface charge density $\sigma$ induced on the surface).
For the bare metal, the surface charge density is mostly located
in a sheet about 1~\Angst past the surface Pt atom.
In the CO-terminated case, this sheet is replaced by a dipolar distribution
on the C atom, followed by most of the surface charge appearing on
the O atom and extending $0.5$~\Angst$ to 1$~\Angst beyond it.
This illustrates that the induced surface charge is migrating to the end
of the carbon monoxide layer, supporting the fact the Pt-CO is insulated
\emph{after}, rather than by, the CO layer.

\begin{figure}
\includegraphics[width=\columnwidth]{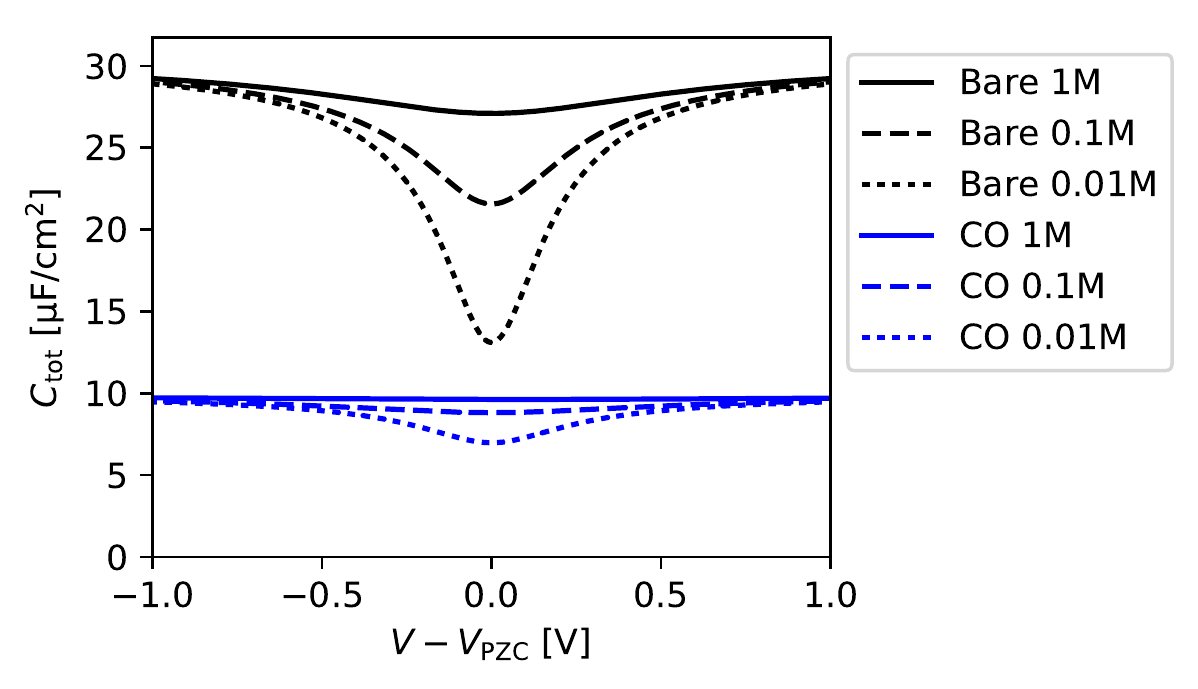}
\caption{Capacitance as a function of electrolyte ionic strength, with the
 Gouy-Chapman-Stern estimate of the diffuse capacitance, in series with the
 $C_m$, $C_a$, and $C_g$ capacitance contributions from Table~\ref{tab:Cap} for bare Pt and Pt-CO.}
\label{fig:CV-GCS}
\end{figure}

The dominance of the gap capacitance is unsurprising considering that the dielectric constant
of the vacuum region is considerably lower than the dielectric constant of the surrounding regions.
Even small gaps between the surface and the solvent result in large drops in potential,
which predominates for hydrophobic surfaces such as the the CO-terminated Pt surface.
Importantly, the induced charge migration to the end of the CO is entirely due to
the electronic structure of the metal electrode and adlayer, and not the solvent.
For this reason, the details of the solvation model do not matter beyond providing
a diffuse capacitance and total capacitance of the overall correct magnitude.
We chose 1 M electrolyte for the solvated DFT calculations because the fluid
capacitance is large and contributes negligibly to the total series capacitance.
Replacing the $C_d\approx 250~\mu$F/cm$^2$ (Table~\ref{tab:Cap}) with the
Gouy-Chapman-Stern model results in the ionic concentration and potential
dependent capacitance prediction shown in Fig.~\ref{fig:CV-GCS}, for potentials
above and below the potential of zero charge of each surface.
The Pt-CO capacitance is expected to exhibit much smaller variations with
ionic concentration and potential because it is dominated by the
gap capacitance which is insensitive to these parameters.  Experimentally, the capacitance variation
with ionic strength is also quite low, and less than the experimental error.  However, this may also be due to 
the potential difference between the potential of zero charge of the Pt-CO ~\cite{cuesta}
and the potential at which the capacitance is measured.

In conclusion, we find the low capacitance of the Pt-CO interface is primarily a result of the gap capacitance
that occurs between the oxygen of the CO, and the solvent.  The CO, strongly electronically-coupled to the Pt, acts as an extension of the metal,
with most of the change in charge with potential occurring at the oxygen atom.  Additionally, the CO capacitance is 
expected to be more insensitive than the bare Pt surface to the ionic strength of the electrolyte, as predicted from
the Gouy-Chapman-Stern diffuse capacitance, coupled with the DFT-predicted  $C_m$, $C_a$, and $C_g$ capacitance terms.

\textbf{Acknowledgements.}
RS acknowledges start-up funding from the Department of Materials
Science and Engineering at Rensselaer Polytechnic Institute.
Calculations were performed on the BlueGene/Q supercomputer in the
Center for Computational Innovations (CCI) at Rensselaer Polytechnic Institute.

\textbf{Supporting Information.} Experimental voltammograms and details
on extracting electrochemical capacitance from these measurements.

\makeatletter{}\providecommand{\latin}[1]{#1}
\makeatletter
\providecommand{\doi}
  {\begingroup\let\do\@makeother\dospecials
  \catcode`\{=1 \catcode`\}=2 \doi@aux}
\providecommand{\doi@aux}[1]{\endgroup\texttt{#1}}
\makeatother
\providecommand*\mcitethebibliography{\thebibliography}
\csname @ifundefined\endcsname{endmcitethebibliography}
  {\let\endmcitethebibliography\endthebibliography}{}

\end{document}